\documentclass{article}
\title
{Gauge  fixing problem and the constrained quantization}
\author{ M.~K.~G\"{u}m\"{u}\c{s}, M.~Boz\\
Hacettepe University, Department of Physics Engineering\\
  06800, Ankara, TURKEY}
\begin{document}
  \maketitle

\begin {abstract}
\noindent In this work, the quantization of  the Yang-Mills theory  is worked out by means of Dirac's canonical quantization method,  using the generalized Coulomb gauge fixing conditions. Following the construction of the matrix composed of all the second class constraints of the theory,  its  convenience  within the framework of the canonical approach is discussed. Although this method can be used successfully in the quantization of  the Abelian theories,  it brings along difficulties for the non-Abelian case, which can not be handled easily even for the generalized Coulomb gauge of the Yang-Mills theory.
\end{abstract}

\section {Introduction}

\noindent  The quantization of gauge fields, which  is necessary for the explanation of the interactions in nature in the framework of the quantum theory,  has been a fundamental backbone for any system at hand, and in the development of the quantization procedure of the gauge theories, various methods have been suggested \cite{Dirac-1}$-$\cite{Feynman-4}. The path integral formalism  is one of those famous methods whose first signal has been appeared in the work of Dirac \cite{Dirac-0}. This formalism was developed by Feynman, as a first step,  for a classical mechanical system  at the end of 40s  \cite{Feynman-1} and accordingly  the mathematical formulation of the quantum theory of electromagnetic interaction was described in the Lagrangian form of quantum mechanics \cite{Feynman-4}.

\noindent  The main reason why  quantization process  plays an important role  in many gauge theories is that 
 they have residual degrees of freedom; thereby posing a serious obstacle in the path of the standard quantization methods of the system. In this context, the constrained systems,  where the degrees of freedom of the fields describing them mathematically are higher than those of determined by observational methods in nature,  has been focus of interest in various works in the literature.

\noindent  From the historical point of view,  the subtlety in the quantization of the constrained systems is first addressed by Dirac in the beginning of 50s,  and it  was again his  success to develop a generalized Hamiltonian dynamics  for the constrained systems  in the context of  the method of canonical quantization \cite{Dirac-11}. Dirac's theory of constrained dynamical systems was  indeed the most general mathematical description of the gauge theories and the progress toward this goal was carried out by Anderson and Bergman whom complete the examination of  the constraints in the Hamiltonian formulation of theories with invariance properties\cite{Anderson-Bergman-14}.

\noindent  On the other hand,  the non-Abelian gauge fields, forming  the basis of quantum chromodynamics,  has been  initiated in the 50s,  with the classical work of Yang and Mills \cite{Yang-Mills-5}, in which  the generalization of the U(1) symmetry of Abelian gauge theory to SU(N) symmetries was carried out  contemporarily, with the developments in the Abelian theory\cite{Dirac-12-58}.  This proposal was followed by the the modeling of strong \cite{ Gell-Mann-9;1961, Zweig-10;1964} and electroweak \cite{Glashow-6, Weinberg-7, Salam-8}    interactions  by using gauge field theories.

\noindent At the end of 60s, the works of Feynman and Dirac has already been well-known methods of quantization \cite {Feynman-4,  Dirac-14} and on one side as  the difficulties of finding  a conventional canonical quantization method for the Yang-Mills theory have  been acquainted with Schwinger's works \cite{Schwinger-1962, Schwinger-1962-2}, on the other side a new formalism was suggested by Fadeev-Popov \cite{Fadeev-Popov-15}  to quantize the gauge fields in the framework of Feynman's path integral formalism. A generalization of the Feynman's method, adapted for the quantization of Yang-Mills field was carried out by Fadeev \cite{Fadeev-16}, with an emphasis on the Hamilton formulation of a constrained system \cite{Mohapatra-1971}. 

\noindent In the light of these developments, it was shown by the pioneer works of t'Hooft, Taylor and Slavnov in 70s \cite{Hooft-18, Taylor-20, Slavnov-19} that the renormalizability and the unitarity  were guaranted by the standard gauge fixing procedure of the general gauge theory formulated by Fadeev-Popov \cite{Fadeev-Popov-15}. The theory, providing a useful platform for the perturbative calculations \cite{Abers-Lee-1973},  later developed by the contribution of many other scientists \cite{Becchi-21, Tyutin-23-1975, Becchi-22}, particularly in the context of the gauge fixing\cite{Hanson-1976}.   
However, at the non-perturbative level,  it was shown by Gribov \cite{Gribov-24-1977} that  the gauge fixing condition does not necessarily lead to a unique potential, in the Coulomb gauge \cite{Tyburski-1978}.

\noindent The gauge fixing process has been permanently an important issue in the quantization of the gauge theories, in the context of getting rid of the redundant degrees of freedom.
However, arising from  the major differences between Abelian and non-Abelian gauge theories,
for instance the existence of self interactions in the non-Abelian case \cite{Abers-Lee-1973},  
although the gauge fixing  is quite straightforward in an Abelian theory,  it becomes a tricky issue for the non-Abelian case. 
In this context, careful treatments of the non-Abelian gauge symmetry
 were extended by various works \cite{Creutz-1978, Jackiw-28-1980, Pak-29-1980, Kihlberg-27-1982} whereas the key ingredients of the Dirac  formalism was reviewed in \cite {Marnelius-17-1982,  Fadeev-1984}. 

\noindent  An economical and a shortcut method in the quantization of the constrained systems was suggested by Fadeev and Jackiw at the end of the 80s \cite{Fadeev-Jackiw-17}. More recent works in the context of the constrained systems \cite{Henneaux:1992} include the relationship between the Dirac and the reduced phase space quantizations \cite{Plyushchay-1993}, 
the generalization the Abelian Coulomb gauge condition to the non-Abelian Yang-Mills theory
\cite{Cranstrom-25}, the gauge fixing processes in the total Hamiltonian formalism\cite{Shirzad-2002}. 

\noindent  An analysis of earlier literature shows  that  the quantization of the non-Abelian theory can be achieved by means of an effective tool such as the Fadeev Popov's device \cite{Fadeev-Popov-15} and it has many advantages.  However,  from the viewpoint of canonical formalism the gauge fixing process is more obscure,  in particular because of the ambiguties in the Coulomb gauge, pointed out by Gribov \cite{Gribov-24-1977}.

\noindent  In this work  the gauge fixing problem for a non-Abelian theory, as an example, that of the Yang-Mills, in the framework of Dirac's constraint formalism is presented. Due to the complexity  of finding a convenient canonical quantization method for the Yang-Mills theory \cite{Schwinger-1962, Schwinger-1962-2, Mohapatra-1971}, Dirac's constraint formalism, being one of the tools to approach this problem,  has  been  widely  used  in  gauge field theories  of  high  energy  physics \cite{Hanson-1976}. Several extension of this method were proposed in recent  studies and one of the latest involves,  for instance,  how the formalism can be applied in curved spacetimes \cite{Cresswell-2015}.  
In the Dirac's  formalism, following the canonical quantization  procedure,  the  total Hamiltonian is composed by  adding  the  constraints  multiplied  by  Lagrange  multipliers  to the  canonical  Hamiltonian; 
\begin{eqnarray}
H_{p}\equiv H_{can}+\lambda_{m}\varphi_{m}\,,\label{1}
\end{eqnarray}
where $\lambda_{m}$'s are Lagrange multipliers; and $\varphi_{m}$'s  are the primary constraints. Thus,  the  action  corresponding  to  this  Hamiltonian  will  contain unknown arbitrary  parameters.  The consistency condition of  the primary constraint,
\begin{eqnarray}
\dot{\varphi}_{m}  \approx  \left\{ \varphi_{m},H_{p}\right\} 
\approx  \left\{ \varphi_{m},H\right\} +C_{mn}\lambda_{n}\approx0\,.\label{2}
\end{eqnarray}
leads to a  new type of constraints, so-called as the  secondary constraints,  if the determinant of the 
elements of the matrix  $C_{mn}=\left\{ \varphi_{m},\varphi_{n}\right\}$ which is constructed by the Poisson parenthesis of the constraints is weakly equivalent to zero. Otherwise,  in the case of the determinant of  $C_{mn}$, doesn't vanish weakly, the Lagrange multipliers, $\lambda_{n}$'s can be determined uniquely. The procedure is repeated until no more constraint or a Lagrange multiplier is obtained. 

\noindent On the other hand, in Dirac's formalism\cite{Dirac-11, Dirac-12-58, Dirac-14}, the constraints are classified as the first and second class.  Due to the fact that the first class constraints satisfy a closed algebra, the system can be quantized only if it has  first class constraints. However, the system should possess  second class constraints to determine the  Lagrange multipliers.

\noindent  Therefore, to get rid of this dilemma for  a constraint system a general method is  necessary in the context of treating both classes on the same basis~\cite{Henneaux:1992}. In this context,  given a constraint system with  second class constraints,  redefining second class constraints and the Hamiltonian by means of an extended phase space for finding an equivalent first class system was one of the possibilities to perform this process,  finally  in which case the second class constraints were transformed into the first class and therefore both of them would have been handled in the same platform. 

\noindent To define this algebra, a matrix which is composed of the Poisson parenthesis of the second class constraints of the system 
\begin{equation}
\Delta_{\alpha\beta}=\left\{ \chi_{\alpha},\chi_{\beta}\right\} .\label{3}
\end{equation}
must be constructed, and to get rid of the redundant degrees of freedom, before quantization;
once the the inverse of this matrix is obtained, the process goes on  by  using the Dirac parenthesis,  instead of the Poisson's. 
This fact is known to be one of the main characteristics of the Dirac formalism, and one of the main motivations for developing the Hamiltonian dynamics of a constrained system, which stems from the the close relation between  the Dirac paranthesis and the quantum commutators \cite{Pak-HU-2015}.

\noindent In the application of this method to any of the gauge systems, two first class constraints are found, after the constraint analysis. To convert these first class constraints to the second class, namely, to reduce the number of degrees of freedom, two gauge fixing conditions should be added to the system “by hand”. These conditions must guarantee the compatibleness with the first class constraints; for instance the transversality property of the gauge fields is provided by the generalized Coulomb gauge, as a pair of generalized Gauss' law constraint, $\left(\mathcal{D}_{i}\Pi^{i}\right)^{a}=0 $.  However, another constraint should have been found to match with the primary constraint,  $\Pi^{a}_{0}=0$. This subtle issue can be overcomed by using the consistency check of the generalized Coulomb gauge. This consistency check  provides the stability of the phase space of the system and also gives a new second class constraint for the system which pairs up with  $\Pi^{a}_{0}=0$.

\noindent The plan of this paper is as follows: Following the brief review of the constraint analysis in the Yang-Mills theory, by means of Dirac canonical quantization method in Section 2,  generalized Coulomb gauge fixing conditions, accompanying the first class constraints in the framework of the theory  are obtained in Section 3. The final  part  involves the construction of  $\Delta_{\alpha\beta}$ (\ref{3})  and its inverse which is necessary at the stage of  passing the Dirac brackets and the discussions on the difficulties yielded within the framework of this approach;  and  thereby the compatibility of the method for this case.

\section {Constraint Analysis in the Yang-Mills Theory} 

\noindent The pure Yang-Mills  theory is described by the Lagrangian density,\footnote[1]{For convention, $c=\hbar=1$ is chosen with the Minkowskian metric (1,-1,-1,-1)~.}
\begin{eqnarray}
\mathcal{L}_{YM}	=-\frac{1}{4}F_{\mu\nu}^{a}F^{\mu\nu a}=\frac{1}{2g^{2}}Tr\left(F_{\mu\nu}F^{\mu\nu}\right),
\label{4}
\end{eqnarray}
where the field tensor $ F^{\mu\nu,a}$ is defined by;
\begin{eqnarray}
F^{\mu\nu,a}=\partial^{\mu}A^{\nu,a}-\partial^{\nu}A^{\mu,a}-gf^{abc}A^{\mu,b}A^{\nu,c}. 
\label{5}
\end{eqnarray}
 Here, $f^{abc}$s are the structure constants of the compact Lie group \footnote[2]{ Being $T^{a}$ is the anti-Hermitean generator of the SU(N) group, the Lie algebra and the normalization of the generators is defined by:
\begin{eqnarray}
 \left[T^{a},T^{b}\right]	=	f^{abc}\, T^{c}~,\nonumber\\
Tr\left(T^{a}T^{b}\right)	=-	1/2 \, \, \delta^{ab}~. 
%\label{1}
\nonumber
\end{eqnarray}} and  the gauge potentials are a set of vector fields denoted by $A_{\mu}^{a}$.

\noindent Quantization of the non-Abelian gauge fields by using the Dirac formalism requires the construction of the the canonical momenta\cite{Pak-29-1980};
\begin{eqnarray}
\Pi^{\mu,a}\left(x\right)\equiv\frac{\partial\mathcal{L}}{\partial\left(\partial_{0}A_{\mu}^{a}\right)}=-F^{0\mu,a}, 
\label{6}
\end{eqnarray}
with the components
\begin{eqnarray}
\Pi^{0,a} &=& 0,\nonumber\\
\Pi^{i,a} &=& E^{i,a},
\label{7}
\end{eqnarray}
where $E^{i,a}$ is the non-Abelian electric field. Using Eq (\ref{6}), one can obtain the equal-time Poisson brackets:
\begin{eqnarray}
\left.\left\{ A_{\mu}^{a}\left(x\right),A_{\nu}^{b}\left(y\right)\right\} \right|_{x_{0}=y_{0}}=0=\left.\left\{ \Pi_{\mu}^{a}\left(x\right),\Pi_{\nu}^{b}\left(y\right)\right\} \right|_{x_{0}=y_{0}},
\label{8}
\end{eqnarray}
\begin{eqnarray}
\left.\left\{ A_{\mu}^{a}\left(x\right),\Pi_{\nu}^{b}\left(y\right)\right\} \right|_{x_{0}=y_{0}}=i\,\delta^{ab}\,\delta_{\mu\nu}\,\delta^{3}\left(\vec{x}-\vec{y}\right).
\label{9}
\end{eqnarray}
In searching the canonical momenta conjugate to the field variables, one encounters with the familiar gauge problem that the momentum conjugate to $A_{0}$ gives no contribution, as seen from Eq (\ref{7}). Therefore, it is expected that the Poisson brackets of \,  $\Pi^{0,a}$ and all the other dynamical variables vanish, which conradicts with  Eq (\ref{9}).
In order to circumvent this obstacle for the  quantization, one makes use of the gauge invariance of the theory  and  incorporate 
 $\Pi^{0,a}=0$ into  the system as a constraint.  The existence of such a constraint necessitates the utilization  of a quantization method developed by the constrained systems, where the canonical variables are the coordinates $A_{i}^{a}$ and their conjugate momenta  $ E_{i}^{a}$.

\noindent Therefore, as a first step,  focusing on the Dirac's canonical quantization method 
the Hamiltonian of the system can be  computed as:
\begin{eqnarray}
\mathcal{H}_{can}& =& \Pi_{\mu}^{a}\partial^{0}A^{\mu,a}-\mathcal{L}\nonumber\\
&=&\Pi_{0}^{a}\,\partial^{0}A^{0,a}+\Pi_{i}^{a}\partial^{0}A^{i,a}-\mathcal{L}\nonumber\\
&=& \Pi_{i}^{a}\big [ -\Pi^{i,a}+\left(\mathcal{D}^{i}A^{0}\right)^{a} \big]-\mathcal{L}\nonumber\\
&=& \mathcal{H}_0 +\Pi_{i}^{a}\left(\mathcal{D}^{i}A^{0}\right)^{a}.
\label{10}
\end{eqnarray}
Here,  the covariant derivative $\mathcal{D}_{\mu}$ is represented in the adjoint representation as;
\begin{eqnarray}
\mathcal{D}_{\mu}\equiv I\partial_{\mu}+\left[A_{\mu},\right.
\label{11}
\end{eqnarray}
and in the first part of the canonical Hamiltonian which is given by;
\begin{eqnarray}
\mathcal{H}_0=\frac{1}{2}\left(\Pi^{i,a}\Pi^{i,a}+B^{i,a}B^{i,a}\right),
\end{eqnarray}
 $B^{i,a} $ represents  the non-Abelian magnetic field: 
\begin{eqnarray}
B^{i,a}=\frac{1}{2}\, \epsilon^{ijk}\, F_{jk}^{a}~.
\label{13}
\end{eqnarray}
Using this expression, one can obtain the canonical Hamiltonian as:
\begin{eqnarray}
 H_{can}	&=&	\int d^{3}x \, \mathcal{H}_{can}\nonumber\\
	            &=& 	H_{0}-\int d^{3} \, A^{0,a}\left(\mathcal{D}^{i}\Pi_{i}\right)^{a},
\label{14}
\end{eqnarray}
where the second term is obtained by dismissing the surface terms after the partial integration. 
\noindent At this point, in the context of the Dirac's formalism, identifying the primary constraint  as;
\begin{eqnarray}
\varphi_{1}^{a}=\Pi^{0,a}=0,
\label{15}
\end{eqnarray}
 and taking into account of the  primary constraints by coupling them to the Lagrange multipliers $ \lambda_{1}$, 
 the primary Hamiltonian can be written as:
\begin{eqnarray}
H_{p} &=& H_{can}+\int d^{3}x\lambda_{1}^{a}\varphi_{1}^{a}\nonumber\\
          &=& H_{0}-\int d^{3}x\left[A^{0,a}\left(\mathcal{D}^{i}\Pi_{i}\right)^{a}-\lambda_{1}^{a}\varphi_{1}^{a}\right].
\label{16}
\end{eqnarray}
Next, to get reasonable results, one has to demand for consistency  that the hypersurface of the constraint in the phase space of the system must not evolve in time. In other words, the time derivative of the primary constraints has to vanish at least weakly. Therefore, the demand for the consistency of the constraint $\varphi_{1}^{a}$  (\ref{15}) requires that: 
\begin{eqnarray}
\left\{ \Pi^{0,a}\left( x \right),H_{p}\right\} \approx0\,. 
\label{17}
\end{eqnarray}
One notes that, in carrying out the Dirac's quantization method, as the weak equation is shown by the symbol ``$ \approx$"  , in accordance with the Dirac's preference, the symbol  $``="$  is used for the strong.

\noindent Taking into account of the distribution properties of the Poisson parenthesis, one can easily determine that the only contribution comes from the second term of Eq (\ref{16}):
\begin{eqnarray}
\left\{ \Pi^{0,a}\left(  x \right),H_{p}\right\}& =&- \int d^{3}y \, \big \{\Pi^{0,a}\left(x \right), \, A^{0,b} \left (x \right)  \big\} \left(\mathcal{D}^{i}\Pi_{i}\left( y \right)\right)^{b}\nonumber\\
                                                                                &=& \int d^{3}y\,\delta^{ab}\, \delta^{3}\left(\vec{x}-\vec{y}\right)\left(\mathcal{D}^{i}\Pi_{i}\left( y \right)\right)^{b}\nonumber\\
&=& \big (\mathcal{D}^{i}\Pi_{i}\left(x \right)\big)^{a}\approx0\,.
\label{18}
\end{eqnarray}
which leads to the Gauss' law. 

\noindent As seen from  Eq (\ref{18}),  the consistency condition for the primary constraint poses  a secondary constraint:
\begin{eqnarray}
\varphi_{2}^{a}=\left(\mathcal{D}_{i}\Pi^{i}\right)^{a}\approx0\,~.
\label{19}
\end{eqnarray}
\noindent Therefore, the secondary Hamiltonian can be obtained as: 
\begin{eqnarray}
H_{S} = H_{can}+\int d^{3}x\lambda_{1}^{a}\varphi_{1}^{a}+\int d^{3}x\lambda_{2}^{a}\varphi_{2}^{a},\nonumber\\
= H_{0}+\int d^{3}x\left[\lambda_{1}^{a}\varphi_{1}^{a}+\left(\lambda_{2}^{a}-A_{0}^{a}\right)\varphi_{2}^{a}\right].
\label{20}
\end{eqnarray}
Here, the secondary constraint is coupled to Lagrange multiplier $ \lambda_{2}$. On the other hand, for the consistency of $\varphi_{2}^{a}$  (\ref{19}), the weakly equivalence condition,
\begin{eqnarray}
\dot{\varphi}_{2}^{a} \approx \left\{ \varphi_{2}^{a},H_{S}\right\}, 
\label{21}
\end{eqnarray}
should be guaranteed. Taking into account of the fact that the only contribution to the right handside of  (\ref{21})  comes from the third term of (\ref{20}), one obtains: 
\begin{eqnarray}
\dot{\varphi}_{2}^{a}=\left\{ \varphi_{2}^{a}\left(\vec{x}\right),\int d^{3}y\left(\lambda_{2}^{b}-A_{0}^{b}\right)\varphi_{2}^{b}\left(\vec{y}\right)\right\} \,.
\label{22}
\end{eqnarray}
\noindent
Furthermore, secondary constraints provide:
 \begin{eqnarray}
\left\{ \varphi_{2}^{a}\left( x \right),\varphi_{2}^{b}\left(y \right)\right\} =f^{abc}\varphi_{2}^{c}\left(x \right)\delta\left(\vec{x}-\vec{y}\right),
\label{23}
\end{eqnarray}
as a result of the SU(N) algebra.
Hence, one obtains:
\begin{eqnarray}
\dot{\varphi}_{2}^{a}\left(x\right)=f^{abc}\, \varphi_{2}^{c}\left(x\right)\big [ \lambda_{2}^{b}\left(x\right)-A_{0}^{b}\left(x\right)\big].
\label{24}
\end{eqnarray}
By identifying,
\begin{eqnarray}
\lambda_{2}^{a}=A_{0}^{a},
\label{25}
\end{eqnarray}
the consistency condition for $ \varphi_{2}^{a}$  is provided and the algorithm stops in which case the secondary Hamiltonian  (\ref{20})  reduces to the following expression: 
\begin{eqnarray}
H_{S}=H_{0}+\int d^{3}x\,\lambda_{1}^{a}\varphi_{1}^{a}. 
\label{26}
\end{eqnarray}
Therefore, the two first  class constraints,(\ref{15}) and (\ref{19}) are  obtained. 
The next step is to convert  all constraints into the second class constraints.

\section{Gauge Fixing}
\noindent In order to proceed in the framework of the Dirac algorithm, equal number of  gauge fixing conditions should be added to these first class constraints  "by hand"  and the whole lump of constraint conditions should be converted into the "second class" constraints. 
In this context, the Coulomb gauge is commonly used, since it satisfies the transversality property of the gauge fields, for instance the electromagnetic field:
\begin{eqnarray}
\chi_{2}^{a}\equiv\partial_{i}A_{i}^{a}\approx0\,.
\label{27}
\end{eqnarray}
It is clear that as  $\chi_{2}^{a}$  couples  with  $\varphi_{2}^{a}$,  it  ensures the required property (namely, the conversion of the first class  into the second class constraint), because of  the transversality property of its structure.  However, the issue  of  how to choose the second gauge condition  demands particular attention in  the non-Abelian case, which doesn't lead to any problem  in the Abelian case, due to it is simplicity.
Therefore, before turning our attention to the non-Abelian case, it will be convinient to discuss the gauge choice,
\begin{eqnarray}
 \chi_{1}=A_{0}\approx0,
\label{28}
\end{eqnarray}
 from the viewpoint of a critical perspective in the Abelian theory~\cite{Gumus-30}.  

\noindent In the Abelian case; the gauge fixing conditions are similar in structure with the primary constraints: 
\begin{eqnarray}
\varphi_{1}&=&\, \Pi_{0}\approx0 \  \, \,  \, \, \rightarrow  \chi_{1}=A_{0}\approx0,\nonumber\\
\varphi_{2}&=&\partial_{i}\Pi_{i}\approx0\ \rightarrow \chi_{2}=\partial_{i} A_{i}\approx0~.
\label{29}
\end{eqnarray}
Here, being  $\varphi_{1}$ and $\varphi_{2}$  the primary and secondary constraints of Maxwell theory, 
$ \chi_{1}$ and   $ \chi_{2}$  are the gauge fixing conditions of the system at hand.

\noindent The main point which shouldn't  be ruled out is 
that;
\begin{eqnarray}
\chi_{2}=\partial_{i}A_{i}\approx0,
\label{30}
\end{eqnarray}
should satisfy the consistency condition: 
\begin{eqnarray}
\dot{\chi}_{2}=\partial_{0}\left(\partial_{i}A_{i}\right)=\partial_{i}\left(\partial_{0}A_{i}\right).
\label{31}
\end{eqnarray}
By using the definition of the  spatial component of canonical momenta,
\begin{eqnarray}
\Pi_{i}=-F_{0i}=-\partial_{0}A_{i}+\left(\partial_{i}A_{0}\right), 
\label{32}
\end{eqnarray}
one obtains:
\begin{eqnarray}
\partial_{0}A_{i}=-\Pi_{i}+\partial_{i}A_{0}\,,
\label{33}
\end{eqnarray}
and replacing  (\ref{33}) into (\ref{31}) gives:
\begin{eqnarray}
\dot{\chi}_{2}	=	\partial_{i}\left(-\Pi_{i}+\partial_{i}A_{0}\right)
	=	-\partial_{i}\Pi_{i}+\vec{\nabla}^{2}A_{0}\,.
\label{34}
\end{eqnarray}
The compatibility of $ \chi_{2}$ with the Gauss law requires that the consistency condition 
\begin{eqnarray}
\dot{\chi}_{2}=\vec{\nabla}^{2}A_{0}\approx0\,,
\label{35}
\end{eqnarray}
should be satisfied and this is possible only for the case;
\begin{eqnarray}
A_{0}\approx0\,.
\label{36}
\end{eqnarray}
which is already the choice of the Abelian theory \cite{Gumus-30}. The reason of carrying out this discussion in such a detail is that the situation is quite different in the non-Abelian case at hand and the choice of the Coulomb gauge can not be done without taking into acount of the subtleties discussed above as in the Abelian theory.

\noindent Therefore, in the light of the above discussion of the Abelian case,  returning back to the non-Abelian theory again, the spatial part of this choice is self-evident since it provides the transversality property: 
\begin{eqnarray}
\chi_{2}^{a}=\partial_{i}A^{i,a}\approx0.
\label{37}
\end{eqnarray}
As a first step, the consistency condition for $\chi_{2}^{a}$ should be checked and to carry out this, one starts from  the definition of the spatial  part of the canonical momenta;
\begin{eqnarray}
\Pi_{i}^{a}=-F_{0i}^{a}=-\partial_{0}A_{i}^{a}+\mathcal{D}_{i}A_{i}^{a}\,, 
\label{38}
\end{eqnarray}
one gets:
\begin{eqnarray}
\dot{\chi}_{2}^{a}	=	\partial_{0}\left(\partial_{i}A_{i}^{a}\right)=\partial_{i}\left(-\Pi_{i}^{a}+\mathcal{D}_{i}A_{0}^{a}\right)
	=	-\partial_{i}\Pi_{i}^{a}+\partial_{i}\mathcal{D}_{i}A_{0}^{a}.
\label{39}
\end{eqnarray}
As a result of the compatibility of the Gauss Law, replacing the consistency condition  (\ref{15})  into  (\ref{34}) gives;
\begin{eqnarray}
\partial_{i}\Pi_{i}^{a}=-\rho^{a}=-\left[A_{i},\Pi_{i}\right]^{a}=-f^{abc}A_{i}^{b}\Pi_{i}^{c}, 
\label{40}
\end{eqnarray}
and from which one obtains: 
\begin{eqnarray}
\dot{\chi}_{2}^{a}=\rho^{a}+\partial_{i}\mathcal{D}_{i}^{ab}A_{0}^{b}\approx0\,. 
\label{41}
\end{eqnarray}
As seen from  (\ref{41}), the consistency of $\chi_{2}^{a}$ excludes the gauge choice $A^{0,a}(x) \approx0$ since this choice  requires  $\rho^{a}=0$, and if this choice was made,  the non-Abelian theory would have been reduced to the Abelian case.

\noindent Therefore, since this choice is dismissed, one should  define a new choice\cite{Gumus-30} which should pair up compatibly with $\varphi_{1}^{a}$;
\begin{eqnarray}
\chi_{1}^{a}=M^{ab}A_{0}^{b}+\rho^{a}\approx0\,,
\label{42}
\end{eqnarray}
where
\begin{eqnarray}
M^{ab}\equiv\partial_{i}\mathcal{D}_{i}^{ab}=-\partial_{i}\partial_{i}+f^{abc}A_{i}^{c}\partial_{i}\,.
\label{43}
\end{eqnarray}
As a result, the generalized Coulomb gauge fixing conditions accompaning the first class constraints in the framework of Yang -Mills theory can be chosen as:
\begin{eqnarray}
\chi_{1}^{a}&=&	M^{ab}A_{0}^{b}+\rho^{a}\approx0,\nonumber\\
\chi_{2}^{a}	&=&	\partial_{i}A^{i,a}\approx0\,.
\label{44}
\end{eqnarray}
One notes that the M matrix given in (\ref{44}) corresponds to the spatial part of the Faddeev - Popov matrix,  $M^{ab}(x)=\left[\partial_{i}\mathcal{D}^{i}\right]^{ab}(x)$;  and within the framework of the Dirac's formalism, it  also plays an important role, particularly  in passing to the Dirac parenthesis of the theory.

\section{The Matrix Constructed by the Second Class Constraints}
\noindent The next step in Dirac algoritma is to calculate the Dirac parentesis which requires the construction of the $\Delta$  matrix, composing of all the second class constraints. Denoting $\varphi_{i}^{a}$ and  $\chi_{i}^{a}$ by the general expression  of $\phi_{i}^{a}$, the constraints of the theory form the following set: 
\begin{eqnarray}
\phi_{1}^{a} & \equiv & \varphi_{1}^{a}=\Pi_{0}^{a}\approx0,\nonumber\\
\phi_{2}^{a} & \equiv & \chi_{1}^{a}=M^{ab}A_{0}^{b}+\rho^{a}\approx0,\nonumber\\
\phi_{3}^{a} & \equiv & \varphi_{2}^{a}=G^{a}=\left(\mathcal{D}_{i}\Pi^{i}\right)^{a}\approx0,\nonumber\\
\phi_{4}^{a} & \equiv & \chi_{2}^{a}=\partial_{i}A^{i,a}\approx0\,.
\label{45}
\end{eqnarray}

\noindent Starting from the general definition of the $\Delta$ matrix; 
\begin{eqnarray}
\Delta_{ij}^{ab}\left(x,y\right)\equiv\left.\left\{ \phi_{i}^{a}\left(x\right),\phi_{j}^{b}\left(y\right)\right\} \right|_{x_{0}=y_{0}}\,,
\label{46}
\end{eqnarray}
its non-zero elements are calculated as:
\begin{eqnarray}
\Delta_{12}^{ab}=\left\{ \phi_{1}^{a},\phi_{2}^{b}\right\} =\left\{ \varphi_{1}^{a}\left(x\right), \chi_{1}^{b}\left(y\right)\right\} _{x_{0}=y_{0}}
=M^{ab}\left( x \right)\delta^{3}\left(\vec{x}-\vec{y}\right)\,,
\nonumber
\label{47-a}
\end{eqnarray}
\begin{eqnarray}
\Delta_{21}^{ab}&=&-\Delta_{12}^{ab}\, ,
%\nonumber
\label{47}
\end{eqnarray}
\begin{eqnarray}
\Delta_{22}^{ab}=\left\{ \phi_{2}^{a},\phi_{2}^{b}\right\} =\left\{ \chi_{1}^{a}\left(x\right),\chi_{1}^{b}\left(y\right)\right\} _{x_{0}=y_{0}}
=f^{abc}\rho^{c}\left( x \right)\delta^{3}\left(\vec{x}-\vec{y}\right)\,,
%\nonumber
\label{48}
\end{eqnarray}
\begin{eqnarray}
\Delta_{23}^{ab}&=&\left\{ \phi_{2}^{a},\phi_{3}^{b}\right\} = \left\{ \chi_{1}^{a}\left(x\right),\varphi_{2}^{b}\left(y \right)\right\} _{x_{0}=y_{0}} = K^{ab}( x, y)  \nonumber\\
	                                                       &=& -f^{abc}\,\bigg[
\varphi_{2}^{c}\left( x \right)\delta^{3}\left(\vec{x}-\vec{y}\right)
		               - 
\partial_{i}\left[A_{0}^{c}\,\partial_{i}\delta^{3}\left(\vec{x}-\vec{y}\right)\right]  \bigg]  \nonumber\\
		            &+&  \bigg[ \delta^{ab}\left(A_{0}^{c}\chi_{2}^{c}\right)-\chi_{2}^{a}A_{0}^{b}\bigg]    \delta^{3} \left(\vec{x}-\vec{y}\right)\,~, \nonumber\\
\nonumber
%\label{49-2}
%&=&K^{ab}(\vec{x},\vec{y})~.
\end{eqnarray}
\begin{eqnarray}
\Delta_{32}^{ab}=-\Delta_{23}^{ab}\,,
%\nonumber
\label{49}
\end{eqnarray}
\begin{eqnarray}
\Delta_{33}^{ab}	=	\left\{ \phi_{3}^{a},\phi_{3}^{b}\right\} =\left\{ \varphi_{2}^{a}\left(x\right),\varphi_{2}^{b}\left(y\right)\right\} _{x_{0}=y_{0}}=f^{abc}G^{c}\left( x\right)\delta^{3}\left(\vec{x}-\vec{y}\right)\,,
\label{50}
%\nonumber
\end{eqnarray}
\begin{eqnarray}
\Delta_{34}^{ab}=	\left\{ \phi_{3}^{a},\phi_{4}^{b}\right\} =\left\{ \varphi_{2}^{a}\left(x\right),\chi_{2}^{b}\left(y\right)\right\} _{x_{0}=y_{0}}=-M^{ab}\left(x \right)\delta^{3}\left(\vec{x}-\vec{y}\right)~,
%\label{50-2}
\nonumber
\end{eqnarray}
\begin{eqnarray}
\Delta_{43}^{ab}=-\Delta_{34}^{ab}~.	
\label{51}
\end{eqnarray}
At this stage, to obtain the Dirac parenthesis, the construction of $\Delta^{-1}$,  shown by  $\Delta^{-1}=\Lambda$, for convenience,  is necessary.  

\noindent  The elements  of $\Lambda$, in terms of those of  $\Delta$  can be calculated as:
\begin{eqnarray}
\Lambda_{11}^{ab}=\left[M^{-1}(x)\right]^{ad}f^{dec}\rho^{c}\left(x\right)\left[M^{-1}(x)\right]^{eb}\delta^{3}\left(\vec{x}-\vec{y}\right)\,,
%\nonumber\
\label{52}
\end{eqnarray}
\begin{eqnarray}
\Lambda_{12}^{ab}=-\left[M^{-1}(x)\right]^{ab}\delta^{3}\left(\vec{x}-\vec{y}\right)\,,
\nonumber
%\label{54}
\end{eqnarray}
\begin{eqnarray}
\Lambda_{21}^{ab}=-\Lambda_{12}^{ab}~,
\label{53}
\end{eqnarray}
\begin{eqnarray}
 \Lambda_{14}^{ab} & = & \left[M^{-1}(x)\right]^{ad}K^{de}(x,y)\left[M^{-1}(x)\right]^{eb}~,
\nonumber
%\label{54}
\end{eqnarray}

\begin{eqnarray}
\Lambda_{41}^{ab}=-\Lambda_{14}^{ab}~, \,\,\,\, \,\,\,
\label{54}
%\nonumber
\end{eqnarray}
\begin{eqnarray}
\Lambda_{34}^{ab}=\left[M^{-1}(x)\right]^{ab}\delta^{3}\left(\vec{x}-\vec{y}\right)\,,
\nonumber
\label{54-1}
\end{eqnarray}
\begin{eqnarray}
\Lambda_{43}^{ab}=-\Lambda_{34}^{ab}~,
%\nonumber
\label{55}
\end{eqnarray}
\begin{eqnarray}
\Lambda_{44}^{ab}=\left[M^{-1}(x)\right]^{ad}f^{dec}G^{c}\left(x\right)\left[M^{-1}(x)\right]^{eb}\delta^{3}\left(\vec{x}-\vec{y}\right)\,.
%\nonumber
\label{56}
\end{eqnarray}
\noindent One notes that  like $\Delta$, the elements of its inverse matrix are also quite complicated
which give rise to technical difficulties  in passing the Dirac parenthesis and moreover this
fact also produces serious suspicions about the utility of this method.

\noindent In any case, taking into account of this fact,  one can  construct the Dirac parenthesis of the theory in terms of the elements of $\Lambda^{ab}$ matrix:
\begin{eqnarray}
\left. \left\{ A_{\mu}^{a}(x),A_{\nu}^{b}(y)\right\} _{D} \right |_{x_{0}=y_{0}} & = & -\delta_{\mu0}\, \delta_{0\nu}\Lambda_{11}^{ab}( x, y)\nonumber\\
 &  & +\big[ \delta_{\mu0}\, \delta_{i\nu} A_{i}^{e}(y)  -\delta_{\mu i}\,\delta_{0\nu} A_{i}^{e}(x)  \big] f^{deb}
\Lambda_{12}^{ad}(x,y)~,\nonumber\\
\label{57}
\end{eqnarray}
\begin{eqnarray}
\left. \left\{ \Pi_{\mu}^{a}(x),\Pi_{\nu}^{b}(y)\right\} _{D}\right |_{x_{0}=y_{0}} &=&
\delta_{\mu i}\, \delta_{j\nu}  \bigg[    \big[  f^{ace}\Pi_{i}^{e}(x)\partial_{j}^{(y)}
+ f^{bce}\Pi_{j}^{e}(y)\partial_{i}^{(x)}    \big]    \Lambda_{34}^{cb}(x,y) \nonumber\\
&+&\partial_{i}^{(x)}\partial_{j}^{(y)}\Lambda_{44}^{ab}(x,y)  \bigg] ~,
\label{58}
\end{eqnarray}

\begin{eqnarray}
\left. \left\{ A_{\mu}^{a}(x),\Pi_{\nu}^{b}(y)\right\} _{D}\right |_{x_{0}=y_{0}} & = & \delta_{\mu\nu}\,\delta^{ab}\delta^{3}(\vec{x}-\vec{y})-\delta_{\mu0}\,\delta_{0\nu}\Lambda_{12}^{ad}(x, y) M^{db}(y)\nonumber\\
& & +\delta_{\mu0} \, \delta_{i\nu}\, f^{deb}\Lambda_{12}^{ad}(x,y)  \big[ \partial_{i}\, A_{i}^{e}(y)+\Pi_{i}^{e}(y) \big] \nonumber\\
 &  & -\delta_{\mu0}\,\delta_{i\nu}\, \partial_{i}^{(y)}\,\Lambda_{14}^{ab}(x,y)
+\delta_{\mu i}\, \delta_{j\nu}\mathcal{D}_{i}^{ca}(\vec{x})\, \partial_{j}^{(y)}\Lambda_{34}^{cb}(x,y)~.\nonumber\\
\label{59}
\end{eqnarray}
\noindent
Among all these elements,  the complexity of even the simplest one,  namely  $\left[M^{-1}(x)\right]^{ab}$, 
brings technical problems,   as will be explained in the following:

\noindent   From the definition  of  $\left[M^{-1}(x)\right]^{ab}$, given in (\ref{37}), its inverse can   be expressed as:
\begin{eqnarray}
\left[ M^{-1}(x) \right]^{ab}=\left(\frac{1}{A+B}\right)^{ab}~,
\label{60}
\end{eqnarray}
where 
\begin{eqnarray}
A^{ab}	=	\delta^{ab}\,\vec{\nabla}^{2},\nonumber\\
B^{ab}	=	f^{abc}A_{i}^{c}\partial_{i}~.
\label{61}
\end{eqnarray}
The inverse of M  can only be computed  explicitely by using  a power series expansion, which is shown symbolically as:
\begin{eqnarray}
\left(\frac{1}{A+B}\right)^{ab}&=&\left(\frac{1}{A}\right)^{ab}-\left(\frac{1}{A}\right)^{ac}B^{cd}\left(\frac{1}{A}\right)^{db}\nonumber\\
&+&\left(\frac{1}{A}\right)^{ac}B^{cd}\left(\frac{1}{A}\right)^{de}B^{ef}\left(\frac{1}{A}\right)^{fb}+\cdots
\label{62}
\end{eqnarray}
However, it can be easily seen that from (\ref{62}) that;
\begin{eqnarray}
\left (A^{-1}\right)^{ab}  =  \frac{\delta^{ab}}{4\pi\left|\vec{x}-\vec{y}\right|}=G^{ab}(x,y)~,
\label{63}
\end{eqnarray}
is the Green's function of  the $\vec{\nabla}^{2}$ operator. On the other hand, the second and the third terms of the expansion can be expressed as:
\begin{eqnarray}
\left(A^{-1}\right)^{ac}B^{cd}\left(A^{-1}\right)^{db}  = \int d^{3} z \, \, {\cal {F}}^{ab}( x, y, z )~;
\nonumber
\end{eqnarray}

\begin{eqnarray}
{\cal {F}}^{ab}( x, y, z )=\frac{1}{4\pi\left|\vec{x}-\vec{y}\right|}\, f^{abc}A_{i}^{c}(z)\, \partial_{i}^{(z)}\frac{1}{4\pi\left|\vec{z}-\vec{y}\right|} ~.
\label{64}
\end{eqnarray}

\begin{eqnarray}
\left(A^{-1}\right)^{ac}B^{cd}\left(A^{-1}\right)^{de}B^{ef}\left(A^{-1}\right)^{fb}
=\int\int d^{3}z \, d^{3}w \, \, {\cal {G}}^{ab}( x,  y, z, w )~; 
\nonumber
\end{eqnarray}

\begin{eqnarray}
{\cal {G}}^{ab}( x, y, z, w )=\frac{1}{4\pi\left|\vec{x}-\vec{z}\right|}  
f^{adc}A_{i}^{c}(w)\, \partial_{i}^{(w)}\frac{1}{4\pi\left|\vec{w}-\vec{z}\right|}f^{dbe}A_{i}^{e}(z)\, \partial_{i}^{(z)}\frac{1}{4\pi\left|\vec{z}-\vec{y}\right|}~.
\nonumber\\
\label{65}
\end{eqnarray}

\noindent It can be easily seen that  the next steps will involve also redundant complexities  and indeed no exact solution can be the obtained for  $M^{-1}$ for this case, since it can only be expressed by a infinite serial expansion which contains the Green's function.

\noindent Taking into account of the available procedures for circumventing such kind of problems, Fadeev-Popov path integral method emanates as an effective tool which brings up solutions to these complexities and this is the reason why the method is adopted as the standart quantization procedure for the residual gauge fields and many other constrained systems\cite{Pak-HU-2014}.

\section{Conclusion}
 The gauge fixing process is one of the most important and subtle issues in the quantization of a gauge theory and to overcome the difficulties in  finding a conventional canonical quantization formalism for the non-Abelian theory, for instance  that of the Yang-Mills,    several procedures  have been developed in the literature. In this work, the problem is attracked by imposing the generalized Coulomb gauge constraints in the framework of the Dirac's method.

\noindent  
In the application of  the Dirac's method to  the non-Abelian  theory,  the construction of  a set of Dirac Brackets which should be  compatible with the constraints plays an important  role  in the sense that  whether one can  use  these brackets  as a basis for a  conventional  canonical quantization method or not.  Certainly,  Dirac's formalism is a useful approach in the development of  a quantization procedure for constrained systems, such as the Fermi-Dirac fields. However, in the context of this work and for the generalized Coulomb gauge,  it is observed  that with the field dependent terms appearing in the Dirac brackets, one can not find an exact solution and thereby  it is extremely difficult  to utilize this scheme as an effective tool.   

\noindent Before concluding, it should be emphasized that, although extremely complicated in the canonical formalism, this process  can be handled easily and straightforwardly from the viewpoint of the functional Path Integral Formalism with the Fadeev Popov method whose mathematical and aesthetical advantages elevated this procedure  to a central position in the area of field quantization.

\section* {Acknowledgements}
This work is dedicated to the honorable memory of  an excellent teacher,  researcher  and  a valuable mentor, Prof. Nam{\i}k Kemal Pak, whom devoted his whole life to knowledge and science. At whatever stage it was, he never gave up to struggle 
for the realization of an ideal of enlightment that endeavoured to transfer all of his knowledge to the greatest number of young people.\\

\noindent The authors would also like to thank Hacettepe University for partial support through the HU-BAP Project 014 A602 006

\end{document}